\documentclass[aps,prd,preprintnumbers,superscriptaddress,nofootinbib,notitlepage,floatfix,10pt,twocolumn,eqsecnum]{revtex4-2}
\usepackage[pdftex]{graphicx}
\usepackage{bm,latexsym,amsmath,amssymb,amsfonts,mathrsfs,mathtools}
\usepackage{color}
\allowdisplaybreaks[1]
\usepackage[pdftex,colorlinks=true,linkcolor=blue,citecolor=cyan,backref=page]{hyperref}
\newcommand*{\D}{\mathrm{d}}
\newcommand*{\mpl}{M_{\mathrm{Pl}}}
\newcommand*{\mixing}{\epsilon_{\mathrm{mix}}}
\newcommand*{\mixingf}{w_{\mathrm{mix}}}
\newcommand*{\affine}{s}

\begin{document}
\title{Geometric optics for gravitational waves in dynamical Chern-Simons gravity:
amplitude birefringence and tensor-scalar mixing on general backgrounds}

%
\author{Tsutomu~Kobayashi}
\email[Email: ]{tsutomu@rikkyo.ac.jp}
\affiliation{Department of Physics, Rikkyo University, Toshima, Tokyo 171-8501, Japan}
%
\begin{abstract}
We study the propagation of high-frequency gravitational waves
in dynamical Chern-Simons gravity within the geometric optics approximation.
In contrast to previous work, we allow for a general background with a nonvanishing scalar field, which enables a covariant treatment of amplitude birefringence.
The background scalar gradient introduces third-derivative terms
into the linearized field equations.
These must be treated as subleading for the linearized system to be free of ghosts,
which yields a covariant upper bound on the frequency.
Tensor-scalar mixing is induced by the Weyl curvature through the Newman-Penrose scalar
$\Psi_0$ along the ray.
When the mixing is efficient, the propagation eigenmodes consist of one luminal,
one subluminal, and one superluminal mode, and a violation of the ghost-free condition leads to a classical instability.
From the transport equations at next-to-leading order,
we show that, when the mixing is negligible,
the relative amplitude of the two helicities depends only on
the scalar-field gradients along the ray at the emission and observation events,
independently of the path in between.
\end{abstract}
\preprint{RUP-26-33}
\maketitle

\section{Introduction}

The detection of gravitational waves~\cite{LIGOScientific:2016aoc}
has opened up new ways of testing general relativity.
Observations of gravitational waves allow us to access the physics of the strong gravity regime
and explore the propagating sector of gravity, providing valuable information about the underlying theory of gravity.
The propagation of gravitational waves is a particularly clean probe of gravity beyond general relativity.
Indeed, the nearly coincident detection of GW170817 and GRB 170817A~\cite{LIGOScientific:2017vwq,LIGOScientific:2017zic} placed
a tight limit on the deviation of the speed of gravitational waves from that of light,
which was used to exclude a broad class of scalar-tensor theories within the Horndeski/beyond-Horndeski family~\cite{Baker:2017hug,Creminelli:2017sry,Sakstein:2017xjx, Ezquiaga:2017ekz}.
The same observation allows us to constrain parity-violating scalar-tensor theories as well~\cite{Nishizawa:2018srh}.

Given the importance of understanding the gravitational wave propagation in modified gravity,
geometric optics for high-frequency gravitational waves in scalar-tensor theories
has been developed in Refs.~\cite{Dalang:2019rke, Garoffolo:2019mna, Ezquiaga:2020dao, Kubota:2022lbn, Menadeo:2024uoq}
(see also~\cite{Goyal:2023uvm} for applications).
The geometric optics approach is useful when one studies the propagation
in backgrounds without symmetry, such as an inhomogeneous cosmological background.
With an additional scalar degree of freedom on top of the tensor modes,
tensor-scalar mode mixing occurs in the course of the propagation through
less symmetric backgrounds (see Ref.~\cite{Dalang:2020eaj} for tensor-scalar mode mixing in Horndeski gravity).
Although a broad class of scalar-tensor theories is in the scope of Refs.~\cite{Dalang:2019rke, Garoffolo:2019mna, Ezquiaga:2020dao, Kubota:2022lbn, Menadeo:2024uoq, Goyal:2023uvm, Dalang:2020eaj},
those with parity-violating interactions have not been addressed explicitly for general backgrounds
(see, however, Ref.~\cite{Ezquiaga:2021ler} for parity-violating effects in the \textit{cosmological} gravitational wave propagation beyond general relativity).

In this paper, we study the propagation of gravitational waves in a parity-violating theory of modified gravity
using the geometric optics approximation.
The specific theory we consider is dynamical Chern--Simons (dCS) gravity~\cite{Jackiw:2003pm, Lue:1998mq, Yunes:2009hc}, in which
the Pontryagin density is coupled to a scalar field.
The theory is well motivated from a high-energy physics viewpoint~\cite{Alexander:2007kv, Alexander:2009tp}.
Various aspects of dCS gravity have been studied over the years
(see Ref.~\cite{Alexander:2009tp} for a review).
Remarkably, gravitational waves in dCS gravity propagate at the speed of light
in a Friedmann--Lema\^{i}tre--Robertson--Walker (FLRW) spacetime~\cite{Yunes:2010yf, Alexander:2017jmt},
which is in sharp contrast to the case of other parity-violating scalar-tensor theories~\cite{Crisostomi:2017ugk, Nishizawa:2018srh}.

In the context of dCS gravity,
geometric optics analysis was applied in Ref.~\cite{Garfinkle:2010zx}
to study the propagation speed of gravitational waves and linear stability
(see Ref.~\cite{Ayzenberg:2013wua} for an extension to quadratic gravity.)
However, Ref.~\cite{Garfinkle:2010zx} assumed backgrounds
on which both the scalar field and the Pontryagin density vanish.
This assumption discards amplitude birefringence,
one of the most characteristic phenomena in dCS gravity~\cite{Lue:1998mq, Alexander:2004wk, Alexander:2007kv}.
The terms responsible for amplitude birefringence
involve higher derivatives and thus profoundly
affect the structure of the field equations and the
well-posedness of the Cauchy problem~\cite{Delsate:2014hba}.
Moreover, only the leading-order equations were considered in Ref.~\cite{Garfinkle:2010zx}, while
the transport equations for the amplitude and polarization,
which arise at next-to-leading order, were not addressed.

In this paper, we extend the analysis of Ref.~\cite{Garfinkle:2010zx}
to general backgrounds with a nonvanishing scalar field.
We derive a covariant ghost-free condition,
identify the Weyl-curvature component responsible for tensor-scalar mixing,
and classify the propagation into three regimes according to the strength of the mixing.
For each regime we derive the dispersion relations and the transport equations,
and show, in particular, that amplitude birefringence is determined
solely by the local scalar-field gradients at the endpoints of the ray.

This paper is organized as follows.
In the next section, we introduce the action for dCS gravity
and briefly review the propagation of gravitational waves
in a cosmological background.
In Sec.~\ref{sec:Lagrangian},
we present the field equations relevant for high-frequency gravitational waves
and scalar-field fluctuation propagating on general backgrounds,
clarifying the approximations we use in this paper.
In Sec.~\ref{sec:LO-eq},
we derive the dispersion relations for the propagating eigenmodes
from the leading-order equations
in the geometric optics approximation.
We then study the transport equations for the amplitude of
each propagating mode, obtained from the next-to-leading-order equations,
in Sec.~\ref{sec:transport}.
Our conclusions are drawn in Sec.~\ref{sec:conclusion}.

Our convention for the Riemann tensor follows that of Wald~\cite{Wald:1984rg}:
$R_{\mu\nu\rho}^{\;\;\;\;\;\;\sigma}=\partial_\nu\Gamma^\sigma_{\mu\rho}-\partial_\mu\Gamma^\sigma_{\nu\rho}+\Gamma^\alpha_{\mu\rho}\Gamma_{\alpha\nu}^\sigma-\Gamma^\alpha_{\nu\rho}\Gamma_{\alpha\mu}^\sigma$.

\section{Dynamical Chern--Simons gravity}\label{sec:dCS-gravity}

The action for dCS gravity is given by~\cite{Jackiw:2003pm, Lue:1998mq, Yunes:2009hc}
\begin{align}
        I=\int\D^4x\sqrt{-g}\left[\mathcal{L}_{\textrm{EH}+\phi}
        +\mathcal{L}_{\textrm{CS}}
        +\mathcal{L}_{\textrm{m}}
        \right],
        \label{eq:action-CS-theory}
\end{align}
with 
\begin{align}
        \mathcal{L}_{\textrm{EH}+\phi}&=\frac{\mpl^2}{2}R
        -\frac{1}{2}g^{\mu\nu}\partial_\mu\phi\partial_\nu\phi-V(\phi),
        \\
        \mathcal{L}_{\textrm{CS}}&=
        \frac{\alpha}{4}\phi 
        R_{\mu\nu}^{\;\;\;\;\alpha\beta}\left(\star R_{\alpha\beta}^{\;\;\;\;\mu\nu}\right),
\end{align}
where
\begin{align}
        \star R_{\alpha\beta}^{\;\;\;\;\mu\nu}\coloneq
        \frac{1}{2}\varepsilon^{\mu\nu\rho\sigma}
        R_{\alpha\beta \rho\sigma},
\end{align}
is the dual of the Riemann tensor,
$\varepsilon^{\mu\nu\rho\sigma}$ is the Levi-Civita tensor,\footnote{The Levi-Civita tensor
is defined by $\varepsilon^{\mu\nu\rho\sigma}\coloneq\epsilon^{\mu\nu\rho\sigma}/\sqrt{-g}$,
where $\epsilon^{\mu\nu\rho\sigma}$ is the totally antisymmetric symbol satisfying $\epsilon^{0123}=+1$.
The sign is different from the convention of Wald.}
$\alpha$ is a CS coupling constant,
and $\mathcal{L}_{\textrm{m}}$ is the matter Lagrangian.
The field equations are of the form
\begin{align}
        &\mpl^2 G_{\mu\nu}+\alpha \mathfrak{C}_{\mu\nu}=T_{\mu\nu}^{(\phi)}+T_{\mu\nu}^{(\textrm{m})},
        \\
        &\Box\phi-V_{,\phi}+\frac{\alpha}{4}
        R_{\mu\nu}^{\;\;\;\;\alpha\beta}\left(\star R_{\alpha\beta}^{\;\;\;\;\mu\nu}\right)
        =0,
\end{align}
where $G_{\mu\nu}$ is the Einstein tensor, $\mathfrak{C}_{\mu\nu}$ comes from $\mathcal{L}_{\textrm{CS}}$,
$T_{\mu\nu}^{(\phi)}$ is the energy-momentum tensor of the scalar field,
and $T_{\mu\nu}^{(\textrm{m})}$ is that of matter.

We study the propagation of the metric and scalar-field perturbations
around a general background.
Before doing so, it is instructive to recap the propagation of cosmological tensor modes
in dCS gravity~\cite{Alexander:2004wk}.
(See Refs.~\cite{Nishizawa:2018srh, Zhao:2019xmm, Ezquiaga:2020dao, Jenks:2023pmk} for
the propagation of cosmological tensor modes in more general parity-violating theories of gravity.)

Let us assume that the background scalar field is nonvanishing and time-dependent, and
consider tensor perturbations around an FLRW background,
\begin{align}
        \D s^2=a^2(\eta)\left[-\D \eta^2+\left(\delta_{ij}+h_{ij}\right)
        \D x^i\D x^j\right]
        ,
\end{align}
where $h_{ij}$ is the transverse-traceless tensor perturbation.
As is well known in standard cosmological perturbation theory,
scalar perturbations including the perturbation of $\phi$ are decoupled from
the tensor sector.
The quadratic action for $h_{ij}$ is given by
\begin{align}
        I^{(2)}&=\frac{\mpl^2}{8}\int\D \eta\D^3x a^2\biggl[
                h_{ij}'h_{ij}'-\partial_k h_{ij}\partial_k h_{ij}
                \notag \\ & \quad 
                +\frac{2\alpha\phi'}{\mpl^2a^2}
                \epsilon_{ijk}\left(
                h_{il}'\partial_j h_{kl}'
                -\partial_m h_{il}\partial_j\partial_mh_{kl}
                \right)
        \biggr],
\end{align}
where a prime denotes differentiation with respect to the conformal time $\eta$ and
$\epsilon_{ijk}$ is the totally antisymmetric symbol with $\epsilon_{123}=1$.
The field equation derived from the quadratic action is given by
\begin{align}
        &\left(a^2h_{ij}'\right)'-a^2h_{ij,kk}
        \notag \\ &
        -\frac{2\alpha}{\mpl^2}\epsilon_{kl(i}
        \left[
                \left(\phi'h_{j)k,l}'\right)'
                -\phi'\partial^2 h_{j)k,l}
        \right]
        =0.
        \label{eq:FE-for-tensor-modes}
\end{align}

We consider a circularly polarized wave with the wave vector $\mathbf{k}$,
\begin{align}
        h_{ij}=h_{\textrm{R,L}}(\eta,|\mathbf{k}|)
        e^{(\textrm{R,L})}_{ij}(\hat{\mathbf{k}})e^{i\mathbf{k}\cdot\mathbf{x}},
\end{align}
where the polarization tensor satisfies
\begin{align}
        \epsilon_{ilk} \hat k_l e_{jk}^{(A)}(\hat{\mathbf{k}})
        =-i\sigma_A e_{ij}^{(A)}(\hat{\mathbf{k}}),
\end{align}
with $\hat{\mathbf{k}}=\mathbf{k}/|\mathbf{k}|$ and $\sigma_A=+1$ ($-1$) for $A=$ R (L).
Then, the field equation~\eqref{eq:FE-for-tensor-modes} reduces to
\begin{align}
        \frac{1}{a^2}\left(a^2\mathcal{A}_A h_A'\right)'
        +|\mathbf{k}|^2 \mathcal{A}_A h_A=0,
\end{align}
with
\begin{align}
        \mathcal{A}_A\coloneq 
        1+\sigma_A \frac{2\alpha\phi'|\mathbf{k}|}{\mpl^2a^2}.
        \label{eq:def-cosmological-A-A}
\end{align}
The polarization-dependent factor $\mathcal{A}_A$ signals amplitude birefringence~\cite{Lue:1998mq, Alexander:2004wk, Alexander:2007kv}.
Observational constraints on amplitude birefringence have been obtained in
Refs.~\cite{Okounkova:2021xjv, Ng:2023jjt, Callister:2023tws}.
Note that this effect arises when the background scalar field is not constant.
It can also be seen that gravitational waves travel at the speed of light in an FLRW universe.

It is important to notice that at sufficiently high frequencies either $\mathcal{A}_{\textrm{R}}$
or $\mathcal{A}_{\textrm{L}}$ flips sign, indicating that one of the modes becomes a ghost.
This places an upper limit of $|\mathbf{k}|$ that one can consider without facing ghost instabilities~\cite{Dyda:2012rj, Yagi:2016jml}:
\begin{align}
        \frac{2|\alpha\phi'|}{\mpl^2 a}\cdot\frac{|\mathbf{k}|}{a}<1.
\end{align}
A similar ghost bound can also be obtained from the analysis of linear perturbations of
a static and spherically symmetric black hole endowed with a nonvanishing background
scalar field~\cite{Motohashi:2011ds}.
In what follows we study high-frequency gravitational waves, but
such an upper bound on the frequency must be imposed and one cannot simply take
the high-$|\mathbf{k}|$ limit.

\section{Effective Lagrangian for high-frequency gravitational waves}\label{sec:Lagrangian}

\subsection{Metric and scalar-field perturbations around general backgrounds}

We derive the quadratic action for the metric perturbations and the scalar-field perturbation
around an arbitrary curved background and an arbitrary scalar-field configuration
described by $g_{\mu\nu}^{(\textrm{bg})}$ and $\phi^{(\textrm{bg})}$.
The perturbed metric and scalar field are written as 
\begin{align}
        g_{\mu\nu}^{(\textrm{tot})}&=g_{\mu\nu}^{(\textrm{bg})}+\mpl^{-1}h_{\mu\nu},
        \label{eq:tot-g}
        \\ 
        \phi^{(\textrm{tot})}&=\phi^{(\textrm{bg})}+\delta\phi,
        \label{eq:tot-phi}
\end{align}
where $h_{\mu\nu}$ is defined so that it has the same mass dimension as $\delta\phi$.
For notational simplicity, hereafter we omit (bg) from the background metric and scalar field.

Substituting $g_{\mu\nu}^{(\textrm{tot})}$ and $\phi^{(\textrm{tot})}$ into
the action~\eqref{eq:action-CS-theory} and expanding it to second order in $h_{\mu\nu}$ and $\delta\phi$,
we obtain the quadratic action
\begin{align}
        I^{(2)}[h_{\mu\nu},\delta\phi]=\int\D^4x\sqrt{-g}
        \left(
                \mathcal{L}_{\textrm{EH}+\phi}^{(2)}+\mathcal{L}_{\textrm{CS}}^{(2)}
        \right),
        \label{eq:I2-schematic}
\end{align}
where the explicit expressions for $\mathcal{L}_{\textrm{EH}+\phi}^{(2)}$
and $\mathcal{L}_{\textrm{CS}}^{(2)}$ are presented in Appendix~\ref{app:derivation-of-quadratic-action}.
Schematically, they are of the form
\begin{align}
        \mathcal{L}_{\textrm{EH}+\phi}^{(2)}&\sim
        (\nabla h)^2+(\nabla\delta\phi)^2
        +\frac{\nabla\phi}{\mpl}h\nabla\delta\phi
        +\dots,
        \label{eq:schematic-L1}
        \\ 
        \alpha^{-1}\mathcal{L}_{\textrm{CS}}^{(2)}&\sim
        \frac{\nabla\phi}{\mpl^2}\varepsilon\nabla h\nabla^2 h
        +\frac{\star R}{\mpl}\nabla\delta\phi\nabla h
        \notag \\ & \quad 
        +\frac{\star R\nabla \phi}{\mpl^2} h\nabla h,
        \label{eq:schematic-L2}
\end{align}
where the ellipsis in Eq.~\eqref{eq:schematic-L1}
denotes terms in which no derivatives act on $h_{\mu\nu}$ or $\delta\phi$
(such as $R hh$).
We will eventually ignore such terms because we are interested in high-frequency waves.
The matter Lagrangian contributes only terms without derivatives acting on $h_{\mu\nu}$ and $\delta\phi$.
Therefore, we can also ignore $\mathcal{L}_{\textrm{m}}^{(2)}$ for our purpose.

\subsection{Field equations for high-frequency gravitational waves}

In the geometric optics approximation~\cite{Isaacson:1968hbi, Misner:1973prb, Maggiore:2007ulw}, we look for a solution of the form
\begin{align}
        q_{\mu\nu}=Q_{\mu\nu}e^{iS},
        \qquad
        \delta\phi = \Phi e^{iS},
        \label{eq:ansatz}
\end{align}
where
\begin{align}
        q_{\mu\nu}\coloneq h_{\mu\nu}-\frac{1}{2}g_{\mu\nu}h.
\end{align}
We assign the same phase $S$ to $q_{\mu\nu}$ and $\delta\phi$,
since we seek the eigenmodes of the coupled system.
Let $\lambdabar$ be the characteristic (reduced) wavelength of $h_{\mu\nu}$ and $\delta\phi$, so that
\begin{align}
        k_\mu\coloneq \partial_\mu S
\end{align}
is of order $\mathcal{O}(\lambdabar^{-1})$.
We introduce the length scales $L_k$ and $L_Q$ over which $k_\mu$, $Q_{\mu\nu}$, and $\Phi$ vary:
\begin{align}
        \nabla k\sim \frac{1}{\lambdabar L_k},\qquad \nabla Q_{\mu\nu}\sim \frac{Q_{\mu\nu}}{L_Q},
        \qquad \nabla\Phi\sim \frac{\Phi}{L_Q}.
\end{align}

Let us introduce the length scales that characterize the background spacetime and scalar field.
Suppose that the length scales associated with the Ricci and Weyl tensors are $L_R$ and $L_W$, respectively:
\begin{align}
        R_{\mu\nu}\sim \frac{1}{L_R^2},\qquad C_{\mu\nu\rho\sigma}\sim \frac{1}{L_W^2}.
\end{align}
The two length scales can be different in general.
For example, for an FLRW background $L_R$ is given by the Hubble radius, while $L_W=\infty$
because the Weyl tensor vanishes.
We also assume that
\begin{align}
        \nabla\phi\sim \frac{\mpl \hat A_\phi}{L_\phi},\qquad \nabla^2\phi\sim \frac{\mpl \hat A_\phi}{L_\phi^2},
\end{align}
where $L_\phi$ is the length scale associated with the gradient of the scalar field and $\hat A_\phi$ is dimensionless.
Then, the gravitational field equations imply that
$\mpl^2R\gtrsim\left(\nabla\phi\right)^2$,
leading to $\hat A_\phi\lesssim L_\phi/L_R$.

In geometric optics, we assume that
\begin{align}
        \epsilon \coloneq \frac{\lambdabar}{L}\ll 1,
\end{align}
where $L$ is the smallest of $L_k$, $L_Q$, $L_R$, $L_W$, and $L_\phi$.\footnote{To
estimate $\epsilon$, the following relation is useful:
\begin{align}
        \epsilon\sim \left(\frac{1\,\text{kHz}}{\lambdabar^{-1}}\right)
        \left(\frac{300\,\text{km}}{L}\right).
\end{align}}
In dCS gravity, one has another ``small'' length scale $\ell_{\textrm{CS}}$ that characterizes the scale of $\alpha$ as
\begin{align}
        \alpha\sim \mpl\ell_{\textrm{CS}}^2.
\end{align}
In terms of $\ell_{\textrm{CS}}$ one can define another small parameter,
\begin{align}
        \mixing\coloneq \frac{\ell_{\textrm{CS}}^2}{L^2_W}\ll 1.
        \label{def:emix}
\end{align}
Note that the denominator is given specifically by the length scale associated with the Weyl tensor.
The reason for this will become clear in the next section.
It will turn out there that
$\mixing$ characterizes the mixing terms between $h_{\mu\nu}$ and $\delta\phi$ in the field equations.
Although $\mixing$ is assumed to be small, it is not necessary to require that $\mixing\sim \epsilon$.
In this paper, we rather consider the following three cases:
(i) $\mixing\ll \epsilon$, (ii) $\mixing\gg \epsilon$,
and (iii) $\mixing\sim \epsilon$.
We work up to next-to-leading-order equations in the $\epsilon$ expansion, and
therefore we require, in case (ii), that $\mixing\ll \epsilon^{1/2}$ in order to
justify our approximation that $\mathcal{O}(\mixing^2)$ terms are ignored.

One can estimate $\mixing$ as
\begin{align}
        \mixing \sim 10^{-1}\left(\frac{\ell_{\textrm{CS}}}{10\,\text{km}}\right)^{2}
        \left(\frac{L_W}{30\, \text{km}}\right)^{-2},
\end{align}
and
the frequency of gravitational waves
in case (ii) must satisfy
\begin{align}
        \lambdabar^{-1}\gg \frac{L_W^2}{\ell_{\textrm{CS}}^2L}>
        \frac{L_W}{\ell_{\textrm{CS}}^2}
        \sim 10
        \left(\frac{\ell_{\textrm{CS}}}{10\,\text{km}}\right)^{-2}
        \left(\frac{L_W}{30\, \text{km}}\right)\,\text{kHz},
\end{align}
where $\ell_{\textrm{CS}}\sim 10$ km is the upper bound from neutron-star observations~\cite{Silva:2020acr}
and $L_W^{-2}\sim (30\,\text{km})^{-2}$ corresponds to the Weyl curvature in the vicinity of neutron stars.
This estimate implies that, for ground-based gravitational-wave detectors, case (i) is most realistic.
The Weyl curvature sourced by cosmological density perturbations is given roughly by
$L_W^{-2}\sim \nabla^2$(lensing potential) $\sim H_0^2\delta\sim \delta(10^{23}\,\text{km})^{-2}$, where $H_0$ is the Hubble parameter and
$\delta$ is the density contrast. This is by far smaller.

To avoid the appearance of a ghost mode as suggested by the analysis of cosmological tensor modes,
we assume that
\begin{align}
        \hat\alpha\coloneq
        \frac{\ell_{\textrm{CS}}^2\hat A_\phi}{L_\phi\lambdabar}
        \lesssim 1.
        \label{eq:avoid-ghost}
\end{align}
More precise, covariant conditions for avoiding
ghosts in general backgrounds will be derived in the next section.

Throughout this paper, we treat dCS gravity as a classical field theory
and restrict ourselves to the frequency range~\eqref{eq:avoid-ghost},
within which the linearized system is free of ghosts.
We note, however, that causality~\cite{Serra:2022pzl, Cassem:2026aib} and positivity~\cite{Xu:2024iao}
arguments suggest that dCS gravity regarded as an EFT
has a cutoff $\Lambda\sim \ell_{\textrm{CS}}^{-1}$.
For $\lambdabar \gtrsim \ell_{\textrm{CS}}$,
one finds $\mixing = \epsilon^2(L/L_W)^2(\ell_{\textrm{CS}}/\lambdabar)^2\lesssim \epsilon^2$,
so that only case (i) is realized.
Cases (ii) and (iii)
probe the structure of the classical theory beyond this cutoff.

Each term in the quadratic action is estimated as
\begin{align}
        (\nabla h)^2\sim (\nabla\delta\phi)^2&\sim \frac{1}{\lambdabar^2},
        \\ 
        \frac{\nabla\phi}{\mpl} h\nabla\delta\phi
        &\lesssim \frac{\epsilon\hat A_\phi}{\lambdabar^2},
        \\ 
        R hh &\lesssim \frac{\epsilon^2}{\lambdabar^2},
        \\ 
        \frac{(\nabla\phi)^2}{\mpl^2} hh &\lesssim \frac{\epsilon^2\hat A_\phi^2}{\lambdabar^2},
\end{align}
and 
\begin{align}
        \frac{\alpha\nabla\phi}{\mpl^2} \varepsilon\nabla h\nabla^2h& \sim 
        \frac{\ell_{\textrm{CS}}^2}{\mpl}\cdot \frac{\mpl \hat A_\phi}{L_\phi}\cdot \frac{1}{\lambdabar^3}
        \sim\frac{\hat\alpha}{\lambdabar^2},
        \label{eq:dhddh}
        \\ 
        \frac{\alpha (\star R)}{\mpl}\nabla\delta\phi\nabla h &\sim \frac{\mixing}{\lambdabar^2},
        \\ 
        \frac{\alpha(\star R)\nabla\phi}{\mpl^2}
        h \nabla h&\lesssim \frac{\hat\alpha\epsilon^2}{\lambdabar^2},
\end{align}
where $h\sim \delta\phi$ is assumed and the dependence of their amplitudes is omitted.
Naively, terms of the form~\eqref{eq:dhddh} scale as $\lambdabar^{-3}$
and would dominate in the geometric optics limit.
However, they are at the same time suppressed by the CS length scale $\ell_{\textrm{CS}}$,
yielding the terms of order $\hat\alpha \lambdabar^{-2}\lesssim \lambdabar^{-2}$.
If one instead picked up the highest derivative terms of the form~\eqref{eq:dhddh}
as leading contributions of order $\mathcal{O}(\lambdabar^{-3})$ in the geometric optics approximation,
the linearized system would exhibit instabilities (at the level of classical theory),
as explained in more detail in the next section.

We will ignore all the terms involving the potential $V$.
In particular, we assume that the mass of the scalar field, $m_\phi=\sqrt{V_{,\phi\phi}}$, is negligible,
$m_\phi^2\ll 1/(\lambdabar L)$.

We now drop the terms that are smaller than $\epsilon/\lambdabar^2$,
and thus arrive at the effective quadratic action,
\begin{align}
        I_{\textrm{eff}}^{(2)}[h_{\mu\nu},\delta\phi]=\int\D^4x\sqrt{-g}\mathcal{L}_{\textrm{eff}}^{(2)},
\end{align}
with
\begin{align}
        \mathcal{L}_{\textrm{eff}}^{(2)}&=
        -\frac{1}{8}\nabla_{\mu}h_{\nu\rho}\nabla^\mu h^{\nu\rho}
        +\frac{1}{8}\nabla_\mu h\nabla^\mu h
        -\frac{1}{4}\nabla_\mu h\nabla_\nu h^{\mu\nu}
        \notag \\ & \quad        
        +\frac{1}{4}\nabla_\mu h^{\mu\nu}\nabla^\rho h_{\rho\nu}
        -\frac{1}{2}(\nabla\delta\phi)^2
        \notag \\ & \quad 
        +\frac{1}{\mpl}\nabla^\mu\phi \left(h_{\mu\nu}-\frac{1}{2}g_{\mu\nu}h\right)\nabla^\nu\delta\phi
        \notag \\ & \quad 
        +\frac{\alpha}{4\mpl^2}\varepsilon^{\mu\nu\rho\sigma}\nabla_\mu\phi
        \nabla_\beta h_{\nu\alpha}\nabla_\rho\nabla^\alpha h_\sigma^\beta
        \notag \\ & \quad 
        -\frac{\alpha}{4\mpl^2}\varepsilon^{\mu\nu\rho\sigma}\nabla_\mu\phi
        \nabla_\beta h_{\nu\alpha}\nabla_\rho\nabla^\beta h_\sigma^\alpha
        \notag \\ & \quad 
        +\frac{\alpha}{\mpl}\left(\star R^{\mu\nu\alpha\beta}\right)
        \nabla_\mu\delta\phi\nabla_\alpha h_{\beta\nu}.
\end{align}

We impose the harmonic gauge condition,
\begin{align}
        \nabla_\mu q^{\mu\nu}=0.
        \label{eq:harmonic-gauge}
\end{align}
Note that there is a residual gauge symmetry:
the harmonic gauge condition is not spoiled by a further gauge transformation $x^\mu\to x^\mu+\xi^\mu$
satisfying $\Box \xi_\mu+R_{\mu\nu}\xi^\nu=\Box \xi_\mu+\mathcal{O}(L_R^{-2})\times\xi_\nu=0$.
Under this gauge condition, the field equations are given by
\begin{align}
        &\Box q_{\mu\nu}+\frac{4}{\mpl}\left(
                \nabla_{(\mu}\phi\nabla_{\nu)}\delta\phi
                -\frac{1}{2}g_{\mu\nu}\nabla_\alpha\phi\nabla^\alpha\delta\phi 
        \right)
        \notag \\ & 
        +\frac{2\alpha}{\mpl^2}\varepsilon_{\alpha\beta\gamma(\mu}\nabla^\alpha\phi
        \nabla^\beta\Box q^{\gamma}_{ \nu)}
        \notag \\ &
        +\frac{2\alpha}{\mpl^2}\varepsilon_{\alpha\beta\gamma(\mu}\nabla_\rho\nabla^\alpha\phi
        \left(
                \nabla^\beta\nabla^\rho q_{\nu)}^\gamma
                -\nabla^\beta\nabla_{\nu)}q^{\gamma\rho}
        \right)
        \notag \\ &
        -\frac{4\alpha}{\mpl}\left(\star C_{\mu \alpha \nu \beta}\right)\nabla^\alpha\nabla^\beta\delta\phi
        =0,
        \label{eq:FE-h}
\end{align}
and
\begin{align}
        \Box\delta\phi 
        -\frac{\alpha}{\mpl}\left(\star C^{\mu\alpha\nu\beta}\right)\nabla_\alpha\nabla_\beta
        q_{\mu\nu}=0,
        \label{eq:FE-phi}
\end{align}
where
\begin{align}
        \star C_{\alpha\beta}^{\;\;\;\;\mu\nu}\coloneq
        \frac{1}{2}\varepsilon^{\mu\nu\rho\sigma}
        C_{\alpha\beta \rho\sigma}.
\end{align}
Up to $\mathcal{O}(\epsilon/\lambdabar^2)$, only the Weyl part of $\star R_{\mu\alpha\nu\beta}$
contributes,
because $\star R_{\mu\alpha\nu\beta}$ appears symmetrized in both $(\mu\nu)$ and $(\alpha\beta)$
and with this symmetrization 
the Ricci part of $\star R_{\mu\alpha\nu\beta}$ is dropped.
The Weyl-dependent terms are responsible for the mode mixing between the tensor and scalar degrees of freedom.
This is why we defined $\mixing$ in terms of the Weyl curvature radius $L_W$ in Eq.~\eqref{def:emix}.


The mode mixing is negligible in the case of $\mixing\ll \epsilon$.
In this case we use the standard expansion,
\begin{align}
        Q_{\mu\nu}&=Q_{\mu\nu}^{(0)}+ \epsilon Q_{\mu\nu}^{(1)}+\dots,
        \\
        \Phi&=\Phi^{(0)}+\epsilon \Phi^{(1)}+\dots,
\end{align}
where $\epsilon$ serves as a bookkeeping parameter.
The same expansion scheme can be used even if one has $\mixing\sim \epsilon$.
However, in the case of $\mixing \gg \epsilon$, we have the small parameter $\epsilon/\mixing\ll 1$,
which is much larger than $\epsilon$.
In this case, the next-to-leading-order correction is of order $\mathcal{O}(\epsilon/\mixing)$:
\begin{align}
        Q_{\mu\nu}&=Q_{\mu\nu}^{(0)}+ \frac{\epsilon}{\mixing} Q_{\mu\nu}^{(1')}+\dots,
        \label{eq:exp-ns1}
        \\
        \Phi&=\Phi^{(0)}+\frac{\epsilon}{\mixing} \Phi^{(1')}+\dots.
        \label{eq:exp-ns2}
\end{align}
(In Sec.~\ref{sec:transport}
we will see that the expansions~\eqref{eq:exp-ns1} and~\eqref{eq:exp-ns2} are indeed consistent with the field equations
for $\mixing \gg \epsilon$.)
We substitute the ansatz~\eqref{eq:ansatz} into the field equations~\eqref{eq:FE-h} and~\eqref{eq:FE-phi},
and collect terms at each order in $\epsilon$.
The leading-order equations are used to determine the eigenmodes and the dispersion relations.
The next-to-leading-order equations give the transport equations for
the leading-order amplitudes $Q_{\mu\nu}^{(0)}$ and $\Phi^{(0)}$.

\section{Leading-order equations}\label{sec:LO-eq}

Let us investigate the leading-order terms, i.e., the $\mathcal{O}(\lambdabar^{-2})$ terms,
in the field equations~\eqref{eq:FE-h} and~\eqref{eq:FE-phi}.
We first keep the $\mathcal{O}(\mixing)$ mixing terms in the leading-order equations.
When $\mixing\gg \epsilon$, the mixing terms are indeed relevant at leading order.
When $\mixing\lesssim\epsilon$, one may simply drop them from the leading-order equations.

Substituting the ansatz~\eqref{eq:ansatz} into Eqs.~\eqref{eq:FE-h} and~\eqref{eq:FE-phi}, we obtain
the following $\mathcal{O}(\lambdabar^{-2})$ equations:
\begin{align}
        &k^2\left(Q_{\mu\nu}^{(0)}+\frac{2\alpha}{\mpl^2}i\varepsilon^{\alpha\beta\gamma}_{\;\;\;\;\;\;(\mu}
        \nabla_\alpha\phi k_\beta Q_{\nu)\gamma}^{(0)}
        \right)
        \notag \\ &
        -\frac{4\alpha}{\mpl}\left(\star C_{\mu \alpha\nu \beta}\right)
        k^\alpha k^\beta\Phi^{(0)}=0,
        \label{eq:leading-cov-Q}
        \\
        &k^2\Phi^{(0)}-\frac{\alpha}{\mpl}\left(\star C^{\mu \alpha\nu \beta}\right)k_\alpha k_\beta Q_{\mu\nu}^{(0)}=0,
        \label{eq:leading-cov-Phi}
\end{align}
with $k^2=k^\mu k_\mu$.
These equations generalize the results of Ref.~\cite{Garfinkle:2010zx} to
backgrounds with a nonvanishing scalar field.

To determine the phase $S$, we introduce a reference null geodesic congruence
generated by the null eikonal equation,
\begin{align}
        g^{\mu\nu}\partial_\mu S_0\partial_\nu S_0=0,
\end{align}
subject to some initial conditions $S_0|_\Sigma$ specified on a three-dimensional spacelike hypersurface $\Sigma$.
Here, $S_0$ can be regarded as the phase of a minimally coupled electromagnetic wave
emitted by the same source as the gravitational waves.
Let
\begin{align}
        l_\mu\coloneq \partial_\mu S_0.
\end{align}
The vector $l$ is then tangent to the null geodesic congruence.
We construct a complex null tetrad $(l,n,m,\bar m)$ associated with the congruence.
The normalization conditions are given by
\begin{align}
        l\cdot n=-1,\qquad m\cdot\bar m = 1,
\end{align}
with all other inner products vanishing.
The tetrad is chosen so that it satisfies $\varepsilon^{\mu\nu\rho\sigma}=-4!\cdot i l^{[\mu} n^\nu m^\rho \bar m^{\sigma ]}$.
We require that $n$, $m$, and $\bar m$ are parallel-propagated along the null geodesic congruence,
\begin{align}
        \nabla_ln=\nabla_l m=\nabla_l\bar m=0,
\end{align}
where $\nabla_l=l^\mu\nabla_\mu$.
Such a parallel-propagated null tetrad has been used in the geometric optics
of electromagnetic and gravitational waves in general relativity~\cite{Dolan:2017zgu}.
Let us then define
\begin{align}
        S_1\coloneq S-S_0,\qquad \delta k_\mu\coloneq \partial_\mu S_1,
\end{align}
so that
\begin{align}
        k^\mu = l^\mu + \delta k^\mu.
\end{align}
Equations~\eqref{eq:leading-cov-Q} and~\eqref{eq:leading-cov-Phi} give $k^2=0$,
if $C_{\mu\alpha\nu\beta}= 0$ (i.e., $\mixing= 0$), and
hence
$k^2\simeq 2l^\mu\delta k_\mu=\mathcal{O}(\mixing/\lambdabar^2)$.\footnote{The ignored $\delta k^\mu\delta k_\mu$
could become large due to a cumulative effect over a distance.
This point is discussed later in this section.}
Since $k^2$ is already of $\mathcal{O}(\mixing/\lambdabar^2)$,
one may replace $k_\beta$ with $l_\beta$ inside the parentheses in the first line of Eq.~\eqref{eq:leading-cov-Q}
as long as we work to first order in $\mixing$.
Furthermore, we may replace
$\left(\star C_{\mu\alpha\nu\beta}\right)k^\alpha k^\beta$
with 
$\left(\star C_{\mu\alpha\nu\beta}\right)l^\alpha l^\beta$
in Eqs.~\eqref{eq:leading-cov-Q} and~\eqref{eq:leading-cov-Phi}.
The eigenvalues $l^\mu\delta k_\mu$ can then be obtained.
Once $l^\mu\delta k_\mu\left(=l^\mu\partial_\mu S_1\right)$ is determined,
one can integrate the equation along each null geodesic with the initial condition $S_1|_\Sigma=0$ to
obtain $S_1$.

Since $\left(\star C_{\mu\alpha\nu\beta}\right)l^\alpha l^\beta l^\mu=0$ and
$\left(\star C_{\mu\alpha\nu\beta}\right)l^\alpha l^\beta g^{\mu\nu}=0$,
we can write
\begin{align}
        \left(\star C_{\mu\alpha\nu\beta}\right)l^\alpha l^\beta
        &= i\bar\Psi_0\, m_\mu m_\nu - i\Psi_0\, \bar m_\mu \bar m_\nu
        \notag \\ & \quad
        - 2i\bar\Psi_1\, l_{(\mu}m_{\nu)} + 2i\Psi_1\, l_{(\mu}\bar m_{\nu)}
        \notag \\ & \quad
        - i\left(\Psi_2-\bar\Psi_2\right) l_\mu l_\nu,
        \label{eq:expression-for-Weyl}
\end{align}
where $\Psi_0\coloneq C_{lmlm}$, $\Psi_1\coloneq C_{lnlm}$, and $\Psi_2\coloneq C_{lm\bar m n}$ are
the Newman--Penrose scalars.
In deriving Eq.~\eqref{eq:expression-for-Weyl} we used the relations
\begin{align}
        \varepsilon_{\alpha\beta\gamma \mu}l^\beta m^\gamma &=-i\left(l_\alpha m_\mu -l_\mu m_\alpha\right),
        \\
        \varepsilon_{\alpha\beta\gamma \mu}l^\beta \bar m^\gamma &=+i\left(l_\alpha \bar m_\mu -l_\mu \bar m_\alpha\right),
        \\
        \varepsilon_{\alpha\beta\gamma \mu}l^\beta n^\gamma &=
        +i\left(m_\alpha \bar m_\mu -m_\mu \bar m_\alpha\right).
\end{align}
The mixing terms can thus be expressed in terms of the Newman--Penrose scalars.

The harmonic gauge condition~\eqref{eq:harmonic-gauge} implies that
$k^\mu Q_{\mu\nu}=\mathcal{O}(\epsilon/\lambdabar)\times Q_{\mu\nu}$,
leading to $l^\mu Q_{\mu\nu}=\mathcal{O}(\text{max}(\epsilon/\lambdabar,\mixing/\lambdabar))\times Q_{\mu\nu}$.
In terms of the null tetrad, we can thus write
\begin{align}
        Q_{\mu\nu}&=Q_{\textrm{R}} m_\mu m_\nu+Q_{\textrm{L}}\bar m_\mu\bar m_\nu
        +Q_{\textrm{T}}m_{(\mu}\bar m_{\nu)}
        \notag \\ & \quad 
        -2Q_{\bar m n}l_{(\mu} m_{\nu)} -2Q_{mn} l_{(\mu}\bar m_{\nu)}+Q_{nn}l_{\mu}l_{\nu}
        \notag \\ & \quad 
        +\mathcal{O}(\text{max}(\epsilon,\mixing)).
\end{align}
The four terms containing $n^\mu$ are suppressed by $\epsilon$ or $\mixing$.
As mentioned earlier, there is a residual gauge symmetry under a coordinate transformation
$x^\mu\to x^\mu+\xi^\mu$ with $\Box\xi^\mu=0$.
This coordinate change can be expressed as
\begin{align}
        \xi_\mu=\left(X l_\mu + Y n_\mu +Z_1 m_\mu + Z_2\bar m_\mu\right)e^{iS_0},
\end{align}
where, in order for the requirement $\Box\xi^\mu=0$ to be satisfied to next-to-leading-order in the $\epsilon$ expansion,
we impose that $2\nabla_l\Xi+\nabla_\mu l^\mu\Xi=0$ ($\Xi=X,Y,Z_1,Z_2$).
For the modes with the null dispersion relation $k^2=0$, one can use this residual gauge symmetry
to transform $Q_{\mu\nu}^{(0)}$ as
\begin{align}
        Q_{\mu\nu}^{(0)}&\to Q_{\mu\nu}^{(0)}-2Ym_{(\mu}\bar m_{\nu)}
        -2Z_1l_{(\mu} m_{\nu)} 
        \notag \\ &\quad
        -2Z_2l_{(\mu}\bar m_{\nu)}-2X l_\mu l_\nu.
\end{align}
Thus, $(Q_{\textrm{T}}^{(0)}, Q_{\bar m n}^{(0)}, Q_{mn}^{(0)}, Q_{nn}^{(0)})$
spans the gauge-dependent sector,
while $Q_{\textrm{R}}^{(0)}$ and $Q_{\textrm{L}}^{(0)}$, as well as $\Phi^{(0)}$,
are unaffected by the gauge transformation.\footnote{The scalar field is transformed by the gauge transformation as
$\delta\phi\to\delta\phi-\xi^\mu\nabla_\mu\phi$, showing that the shift is suppressed by $\epsilon$.}
In contrast, for the modes with a dispersion relation $k^2\neq 0$,
there is no residual gauge symmetry to change $(Q_{\textrm{T}}^{(0)}, Q_{\bar m n}^{(0)}, Q_{mn}^{(0)}, Q_{nn}^{(0)})$.
In any case, one can explicitly verify that
only $Q_{\textrm{R}}^{(0)}$ and $Q_{\textrm{L}}^{(0)}$ appear
in the linearized Riemann tensor at leading order in the $\epsilon$ expansion,
\begin{align}
        \mpl\delta R_{\mu\nu\rho\sigma}=
                l_{\sigma} l_{[\mu}q_{\nu]\rho}^{\textrm{phys}}
                -l_\rho l_{[\mu} q_{\nu]\sigma}^{\textrm{phys}}
                +\mathcal{O}(\epsilon,\mixing)
        ,
\end{align}
where
$q_{\mu\nu}^{\textrm{phys}}\coloneq \left(Q_{\textrm{R}}^{(0)} m_\mu m_\nu+Q_{\textrm{L}}^{(0)}\bar m_\mu \bar m_\nu\right)e^{iS}$.
As we will see shortly,
we can derive a closed set of equations for the ``physical'' sector
$(Q_{\textrm{R}}^{(0)}, Q_{\textrm{L}}^{(0)}, \Phi^{(0)})$.

Contracting Eq.~\eqref{eq:leading-cov-Q} with $\bar m^\mu\bar m^\nu$ and $m^\mu m^\nu$,
we obtain
\begin{align}
        l^\mu\delta k_\mu \mathcal{A}_{\textrm{R}} Q_{\textrm{R}}^{(0)}
        -2i\mixingf e^{-i\chi}\Phi^{(0)}
        &=0,
        \label{eq:leading-R}
        \\
        l^\mu\delta k_\mu\mathcal{A}_{\textrm{L}}Q_{\textrm{L}}^{(0)}
        +2i\mixingf e^{i\chi}\Phi^{(0)}
        &=0,
        \label{eq:leading-L}
\end{align}
where $\chi$ is the argument of $\Psi_0$ (i.e., $\Psi_0=|\Psi_0|e^{i\chi}$),
\begin{align}
        \mixingf \coloneq \frac{\alpha|\Psi_0|}{\mpl},
\end{align}
and
\begin{align}
        \mathcal{A}_{\textrm{R}}\coloneq 1+\frac{2\alpha\nabla_l\phi}{\mpl^2},
        \qquad 
        \mathcal{A}_{\textrm{L}}\coloneq 1-\frac{2\alpha\nabla_l\phi}{\mpl^2}.
\end{align}
Equation~\eqref{eq:leading-cov-Phi} now reads
\begin{align}
        l^\mu\delta k_\mu \Phi^{(0)}
        +\frac{i}{2}\mixingf \left(e^{i\chi}Q_{\textrm{R}}^{(0)}
        -e^{-i\chi}Q_{\textrm{L}}^{(0)}\right)
        =0,
        \label{eq:leading-0}
\end{align}
Note that $\lambdabar^2\mixingf\lesssim\mixing$.
Thus, the mode mixing is governed solely by $\Psi_0$.
Hereafter we use $\mixingf$ to quantify the mode-mixing effects.
Accordingly, we redefine cases (i)--(iii) locally by comparing $\lambdabar^2\mixingf$ with $\epsilon$.
(In contrast to $\mixing$, $\mixingf$ is a local quantity that can vary along the null geodesic.)
Note also that $\mathcal{A}_{\textrm{R}}$ and $\mathcal{A}_{\textrm{L}}$ are
the generalization of Eq.~\eqref{eq:def-cosmological-A-A}.
These factors are responsible for amplitude birefringence,
which manifests itself when $\nabla_l\phi\neq 0$.
Recall that $\alpha\nabla_l\phi/\mpl^2=\mathcal{O}(\hat\alpha)$.
We assume that
\begin{align}
        \mathcal{A}_{\textrm{R}}>0,
        \qquad 
        \mathcal{A}_{\textrm{L}}>0,
\end{align}
namely,
\begin{align}
        \frac{2}{\mpl^2}\left| \alpha\nabla_l\phi \right|<1.
        \label{eq:avoid-ghost-cov}
\end{align}
This is the covariant expression for the ghost bound,
specifying the frequency range~\eqref{eq:avoid-ghost} more precisely.

Let us first consider the case of $\lambdabar^2\mixingf\lesssim \epsilon$.
Dropping the terms proportional to $\mixingf$ from
Eqs.~\eqref{eq:leading-R}, \eqref{eq:leading-L}, and~\eqref{eq:leading-0},
we see that the natural eigenmodes are $Q_{\textrm{R}}^{(0)}$, $Q_{\textrm{L}}^{(0)}$,
and $\Phi^{(0)}$, and
they share the null dispersion relation $k^2=0$ in this regime.
This in particular applies to the limiting case of $\mixingf =0$,
which includes backgrounds with vanishing Weyl tensor (i.e., Petrov type O spacetime)
such as FLRW spacetime~\cite{Garfinkle:2010zx}.
Even if the Weyl tensor itself is nonvanishing, one has $\mixingf=0$
if $l$ is a principal null direction.
Radial null geodesics in Schwarzschild spacetime fall into this category, which explains why
the radial propagation speed of gravitational waves is unaffected by the CS term as found in Ref.~\cite{Garfinkle:2010zx}.

The transport equations are derived from
the next-to-leading-order terms (i.e., the $\mathcal{O}(\epsilon/\lambdabar^{2})$ terms) in the field equations.
In the case of $\lambdabar^2\mixingf\sim \epsilon$, the mixing terms must
therefore be removed from the leading-order equations and be absorbed into the transport equations.
In the case of $\lambdabar^2\mixingf\ll \epsilon$, the mixing terms are completely removed
in the analysis up to $\mathcal{O}(\epsilon/\lambdabar^2)$.

Next, let us study the case of $\lambdabar^2\mixingf \gg \epsilon$,
where the mixing terms do come into play in the leading-order equations.
In this case we get the eigenvalues
\begin{align}
        l^\mu\delta k_\mu =0
        \qquad \text{and}\qquad l^\mu \delta k_\mu=
        \pm \delta_W,
        \label{eq:eigenvalues}
\end{align}
where
\begin{align}
        \delta_W\coloneq
        \mixingf
        \sqrt{\frac{2}{\mathcal{A}_{\textrm{R}} \mathcal{A}_{\textrm{L}}}}.
\end{align}
It is clear from Eq.~\eqref{eq:eigenvalues} that there is one luminal, one subluminal, and one superluminal mode.
For the latter two modes, the rays thus deviate from null geodesics.
This result generalizes the finding in Ref.~\cite{Garfinkle:2010zx} to more general backgrounds.
In particular, the nonvanishing background scalar field does not change the conclusion
of the presence of the subluminal and superluminal modes.


Now one can see that the assumption~\eqref{eq:avoid-ghost-cov} is not merely technical:
if the condition is violated, one has $\mathcal{A}_{\textrm{R}}\mathcal{A}_{\textrm{L}}<0$
and $\delta_W$ becomes imaginary,
signaling an instability caused (at the level of classical field theory)
by the mixing between the ghost-like tensor mode
and the scalar fluctuation through the Weyl curvature.

For the mode with $l^\mu\delta k_\mu = 0$ (the luminal mode), one has
\begin{align}
        \Phi^{(0)}=0,\qquad e^{i\chi} Q_{\textrm{R}}^{(0)}=e^{-i\chi} Q_{\textrm{L}}^{(0)}.
\end{align}
Defining the amplitude $\widetilde{Q}_0^{(0)}$ by
\begin{align}
        Q_{\textrm{R}}^{(0)}=\frac{e^{-i\chi}}{\sqrt{2}}\widetilde{Q}_0^{(0)},
        \qquad 
        Q_{\textrm{L}}^{(0)}=\frac{e^{i\chi}}{\sqrt{2}}\widetilde{Q}_0^{(0)},
\end{align}
one can write the eigenmode as
\begin{align}
        q_{\mu\nu}^{\textrm{phys}}&=
        \widetilde{Q}_0^{(0)}e^{iS_0}\mathsf{e}_{\mu\nu}^+,
        \\ 
        \delta\phi&=0,
\end{align}
where
\begin{align}
        \mathsf{e}_{\mu\nu}^+\coloneq \frac{1}{\sqrt{2}}
        \left(e^{-i\chi}m_\mu m_\nu+e^{i\chi}\bar m_\mu\bar m_\nu\right)
        \label{eq:def-plus-rot}
\end{align}
is the ``plus'' polarization tensor rotated by the angle $\chi/2$ in the plane relative to the
parallel-propagated frame.
Similarly, it is convenient to introduce the ``cross'' counterpart,
\begin{align}
        \mathsf{e}_{\mu\nu}^\times\coloneq \frac{i}{\sqrt{2}}
        \left(e^{-i\chi}m_\mu m_\nu-e^{i\chi}\bar m_\mu\bar m_\nu\right),
\end{align}
satisfying $\mathsf{e}^{+\mu\nu} \mathsf{e}_{\mu\nu}^\times = 0$.
The rotation of the polarization plane is contrasted with the case of general relativity,
where the polarization tensor is parallel-propagated along null geodesics.
In the present case, one instead has
$\nabla_l\mathsf{e}_{\mu\nu}^+=-(\nabla_l\chi)\mathsf{e}_{\mu\nu}^\times$
and
$\nabla_l\mathsf{e}_{\mu\nu}^\times=(\nabla_l\chi)\mathsf{e}_{\mu\nu}^+$.
This eigenmode is consists purely of metric perturbations and is linearly polarized.

For the modes with $l^\mu \delta k_\mu = \pm \delta_W$, one has
\begin{align}
        Q_{\textrm{R}}^{(0)}&=
        \pm ie^{-i\chi}\sqrt{\frac{2\mathcal{A}_{\textrm{L}}}{\mathcal{A}_{\textrm{R}}}}
        \Phi^{(0)},
        \\ 
        Q_{\textrm{L}}^{(0)}&=
        \mp ie^{i\chi}\sqrt{\frac{2\mathcal{A}_{\textrm{R}}}{\mathcal{A}_{\textrm{L}}}}
        \Phi^{(0)}.
\end{align}
Defining the amplitudes $\widetilde{Q}_\pm^{(0)}$ by
\begin{align}
        Q_{\textrm{R}}^{(0)}&=ie^{-i\chi}
        \sqrt{\frac{\mathcal{A}_{\textrm{L}}}{2\mathcal{A}_{\textrm{R}}}}
        \widetilde Q_\pm^{(0)},
        \label{eq:def-amp-1}
        \\ 
        Q_{\textrm{L}}^{(0)}&=-i e^{i\chi}
        \sqrt{\frac{\mathcal{A}_{\textrm{R}}}{2\mathcal{A}_{\textrm{L}}}}
        \widetilde Q_\pm^{(0)},
        \\ 
        \Phi^{(0)} &=\pm \frac{1}{2}
        \widetilde{Q}_\pm^{(0)},
        \label{eq:def-amp-3}
\end{align}
one can write the eigenmodes as
\begin{align}
        q_{\mu\nu}^{\textrm{phys}}&=
        \frac{1}{\sqrt{\mathcal{A}_{\textrm{R}}\mathcal{A}_{\textrm{L}}}}
        \widetilde{Q}_\pm^{(0)}e^{i(S_0+S_{1\pm})}
        \notag \\ & \quad 
        \times
        \left[\mathsf{e}_{\mu\nu}^\times
        +i\left(\frac{\mathcal{A}_{\textrm{L}}-\mathcal{A}_{\textrm{R}}}{2}\right)\mathsf{e}_{\mu\nu}^+
        \right],
        \\ 
        \delta\phi&=\pm \frac{1}{2}\widetilde{Q}_\pm^{(0)}e^{i(S_0+S_{1\pm})},
\end{align}
where
\begin{align}
        S_{1\pm}=\pm \int^\affine_{\affine_{\textrm{i}}}\delta_W(\affine')\D\affine',
\end{align}
with $\affine_{\textrm{i}}$ being the initial value of the affine parameter $\affine$ of the reference null geodesic congruence.
These eigenstates are elliptically polarized, with an axial ratio
$2/|\mathcal{A}_{\textrm{L}}-\mathcal{A}_{\textrm{R}}|=\mathcal{O}(\hat\alpha^{-1})$.
Only when $\nabla_l\phi=0$ do we have $\mathcal{A}_{\textrm{R}}=\mathcal{A}_{\textrm{L}}$,
in which case these states are linearly polarized.

We have thus identified
the three modes characterized by the amplitudes
$(\widetilde{Q}_0^{(0)}, \widetilde{Q}_+^{(0)}, \widetilde{Q}_-^{(0)})$ in the $\lambdabar^2\mixingf\gg \epsilon$ regime.
These are analogous to the propagation eigenstates introduced in Ref.~\cite{Ezquiaga:2020dao}.

We are now in a position to discuss cumulative effects.
Consider waves propagating over a distance $D$.
The corresponding affine distance $\Delta \affine$ is given by $\Delta s\sim \lambdabar D$
because $l^\mu \sim D/\Delta s \sim 1/\lambdabar$.
Thus, $S_{1\pm}\sim (\mixing/\lambdabar^2)\cdot \Delta\affine\sim \mixing D/\lambdabar$,
and the transverse component $\delta k_\perp$ of $\delta k^\mu$ is estimated as
$\delta k_\perp \sim S_{1\pm}/L\sim \mixing D/(\lambdabar L)$,
which accumulates in contrast to $l^\mu\delta k_\mu$.
We have ignored $\mathcal{O}(\delta k_\perp^2)$ terms in $k^2$,
but the ignored terms induce an $\mathcal{O}(\delta k_\perp^2\cdot\Delta\affine)$ correction to the phase.
Requiring that this correction is smaller than $\mathcal{O}(1)$,
we obtain $D\lesssim L(\epsilon/\mixing^2)^{1/3}$.
Since we are assuming that $\mixing\ll\epsilon^{1/2}$,
the cumulative effect does not spoil our approximation over a distance much larger than $L$.
This should be contrasted with the limit on the distance over which the geometric optics is valid:
$D\lesssim L/\epsilon$.
(In the geometric optics approximation we neglect $\mathcal{O}(\epsilon^2/\lambdabar^2)$ terms in the field equations,
which is justified as long as $\epsilon^2/\lambdabar^2\cdot\Delta \affine \lesssim 1$.)
Since one has $\mixing\gg\epsilon^2$ in case (ii), the former gives a
stricter bound on $D$ than the latter.

At this stage it is instructive to look into the field equations
for the remaining components of $Q_{\mu\nu}^{(0)}$, which read
\begin{align}
        &l^\mu\delta k_\mu Q_{\textrm{T}}^{(0)}=0,
        \label{eq:rem-T}
        \\
        &l^\mu\delta k_\mu \left[
                \left(1+\frac{\alpha}{\mpl^2}\nabla_l\phi\right)Q_{\bar m n}^{(0)}
                +\frac{\alpha}{\mpl^2}
                \nabla_m\phi Q_{\textrm{R}}^{(0)}
        \right]
        \notag \\ & 
        -\frac{2\alpha}{\mpl}i\bar\Psi_1\Phi^{(0)}
        =0,
        \label{eq:rem-bar-mn}
        \\ 
        &l^\mu\delta k_\mu \left[
                \left(1-\frac{\alpha}{\mpl^2}\nabla_l\phi\right)Q_{m n}^{(0)}-\frac{\alpha}{\mpl^2}
                \nabla_{\bar m}\phi Q_{\textrm{L}}^{(0)}
        \right]
        \notag \\ &
        +\frac{2\alpha}{\mpl}i \Psi_1\Phi^{(0)}
        =0,
        \label{eq:rem-mn}
        \\ 
        &l^\mu\delta k_\mu\left[
                Q_{nn}^{(0)}-\frac{2\alpha}{\mpl^2}
                \left(\nabla_{\bar m}\phi Q_{m n}^{(0)}-\nabla_m\phi Q_{\bar m n}^{(0)}\right)
        \right]
        \notag \\ &
        +\frac{2\alpha}{\mpl}i\left(\Psi_2-\bar\Psi_2\right)\Phi^{(0)}=0,
        \label{eq:rem-nn}
\end{align}
where using Eq.~\eqref{eq:rem-T} we dropped the terms proportional to $l^\mu\delta k_\mu Q_{\textrm{T}}^{(0)}$
from Eqs.~\eqref{eq:rem-bar-mn} and~\eqref{eq:rem-mn}.
For the modes with $l^\mu\delta k_\mu=\pm\delta_W$ in the regime $\lambdabar^2\mixingf\gg \epsilon$,
these equations allow us to express
$(Q_{\textrm{T}}^{(0)}, Q_{\bar m n}^{(0)}, Q_{mn}^{(0)}, Q_{nn}^{(0)})$ in terms
of $(Q_{\textrm{R}}^{(0)}, Q_{\textrm{L}}^{(0)}, \Phi^{(0)})$
(and, in particular, we find that $Q_{\textrm{T}}^{(0)}=0$).
In contrast, setting $l^\mu\delta k_\mu=0$ and $\alpha\cdot$ (Weyl) $\cdot\Phi^{(0)}=0$
makes these equations trivial and $Q_{\textrm{T}}^{(0)}, Q_{\bar m n}^{(0)}, Q_{mn}^{(0)}$, and $Q_{nn}^{(0)}$
are left undetermined.
This reflects the fact that they are gauge dependent in the case of $l^\mu\delta k_\mu=0$.
In any case, we have a closed set of equations for $(Q_{\textrm{R}}^{(0)}, Q_{\textrm{L}}^{(0)}, \Phi^{(0)})$
and this physical sector is not affected by these remaining variables.

\section{Transport equations}\label{sec:transport}

We now turn to the investigation of the transport equations for the
leading-order amplitudes, which are
obtained at next-to-leading order in the $\epsilon$ expansion.

\subsection{\texorpdfstring{Case (i): $\lambdabar^2\mixingf\ll \epsilon$}{Case (i)}}

The analysis in this regime is simple.
Dropping the mixing terms, we obtain, at $\mathcal{O}(\epsilon/\lambdabar^2)$,
\begin{align}
        \mathcal{E}_{\mu\nu}=0,
\end{align}
with
\begin{align}
        \mathcal{E}_{\mu\nu}&\coloneq
        i\left(\nabla_\alpha l^\alpha Q_{\mu\nu}^{(0)}+2l^\alpha\nabla_\alpha Q_{\mu\nu}^{(0)}\right)
        \notag \\ &\quad
        -\frac{2\alpha}{\mpl^2}\varepsilon^{\alpha\beta\gamma}_{\;\;\;\;\;\;(\mu}\nabla_\alpha\phi
        l_\beta
        \left(
                \nabla_\rho l^\rho Q_{\nu)\gamma}^{(0)}+2l^\rho\nabla_\rho Q_{\nu)\gamma}^{(0)}
        \right)
        \notag \\ &\quad
        -\frac{2\alpha}{\mpl^2}\varepsilon^{\alpha\beta\gamma}_{\;\;\;\;\;\;(\mu}l_\beta
        \nabla^\rho\nabla_\alpha\phi
        \left(l_\rho Q_{\nu) \gamma}^{(0)}-l_{\nu)}Q_{\gamma\rho}^{(0)}\right)
        \notag \\ &\quad 
        +\frac{4}{\mpl}\left(l_{(\mu}\nabla_{\nu)}\phi-\frac{1}{2}g_{\mu\nu}\nabla_l\phi\right)\Phi^{(0)}.
        \label{def:emn}
\end{align}
Contracting this with $\bar m^\mu \bar m^\nu$ and $m^\mu m^\nu$
yields the equations for $Q_{\textrm{R}}^{(0)}$ and $Q_{\textrm{L}}^{(0)}$,
\begin{align}
        \mathcal{E}_{\textrm{R}}&\coloneq
        \mathcal{A}_{\textrm{R}}
        \nabla_lQ_{\textrm{R}}^{(0)}
        +\frac{1}{2}\nabla_\mu\left(\mathcal{A}_{\textrm{R}}l^\mu\right)
        Q_{\textrm{R}}^{(0)}
        =0,
        \\
        \mathcal{E}_{\textrm{L}}&\coloneq 
        \mathcal{A}_{\textrm{L}}
        \nabla_lQ_{\textrm{L}}^{(0)}
        +\frac{1}{2}\nabla_\mu\left(\mathcal{A}_{\textrm{L}}l^\mu\right)
        Q_{\textrm{L}}^{(0)}
        =0.
\end{align}
These equations can be recast into the form of conservation equations as
\begin{align}
        \nabla_\mu\left(\mathcal{A}_{\textrm{R}}|Q_{\textrm{R}}^{(0)}|^2l^\mu\right)&=0,
        \label{eq:cons-R}
        \\
        \nabla_\mu\left(\mathcal{A}_{\textrm{L}}|Q_{\textrm{L}}^{(0)}|^2l^\mu\right)&=0.
        \label{eq:cons-L}
\end{align}
The conserved currents contain the factors $\mathcal{A}_{\textrm{R}}$ and $\mathcal{A}_{\textrm{L}}$,
which are positive thanks to the ghost bound.

Consider gravitational waves emitted at an event $e$ and observed at an event $o$
on the same null ray. In case (i), Eqs.~\eqref{eq:cons-R} and~\eqref{eq:cons-L}
imply that $\mathcal{A}_{\mathrm{R}}|Q_{\mathrm{R}}^{(0)}|^2/\mathcal{A}_{\mathrm{L}}|Q_{\mathrm{L}}^{(0)}|^2$
is constant along each ray, so that
\begin{align}
        \left.\left|\frac{Q_{\mathrm{R}}^{(0)}}{Q_{\mathrm{L}}^{(0)}}\right|^2\right|_{o}
        =
        \left.\frac{\mathcal{A}_{\mathrm{L}}}{\mathcal{A}_{\mathrm{R}}}\right|_{o}
        \left.\frac{\mathcal{A}_{\mathrm{R}}}{\mathcal{A}_{\mathrm{L}}}\right|_{e}
        \left.\left|\frac{Q_{\mathrm{R}}^{(0)}}{Q_{\mathrm{L}}^{(0)}}\right|^2\right|_{e}.
        \label{eq:chirality-ratio}
\end{align}
The change in the relative amplitude of the two helicities is thus determined solely by
the values of $\nabla_l\phi$ at the endpoints and is independent
of the path in between.
For an observer with four-velocity $u^\mu$, writing
$l^\mu=\omega(u^\mu+\hat n^\mu)$ gives $\nabla_l\phi=\omega(u^\mu\nabla_\mu\phi+\hat n^\mu\nabla_\mu\phi)$,
so that the relevant quantity involves the locally measured frequency and the gradient
of $\phi$ along the propagation direction $\hat n^\mu$ as well as its time derivative.
Equation~\eqref{eq:chirality-ratio} is a covariant generalization of amplitude birefringence
in cosmology~\cite{Alexander:2007kv, Yunes:2010yf}, where the effect is often expressed as an integral along the
line of sight that in fact reduces to boundary terms.
In cosmology, the growth of the effect with distance
arises entirely from the blueshift of the emitted frequency,
$\omega_e=(1+z)\omega_o$, entering $\nabla_l\phi$ at the source,
rather than from accumulation along the path.
We stress that this result relies
on the separate conservation of the two currents and therefore ceases to hold once
the mode mixing becomes effective.

From the scalar-field equation at $\mathcal{O}(\epsilon/\lambdabar^2)$, we obtain
\begin{align}
        \mathcal{E}_\Phi=0,
\end{align}
with
\begin{align}
        \mathcal{E}_\Phi\coloneq \nabla_l\Phi^{(0)}+\frac{1}{2}\nabla_\mu l^\mu \Phi^{(0)},
        \label{def:ephi}
\end{align}
which also leads to the conservation equation
\begin{align}
        \nabla_\mu\left(|\Phi^{(0)}|^2l^\mu\right)=0.
\end{align}

We have thus derived the three separate conservation equations for the eigenmodes
$Q_{\textrm{R}}^{(0)}$, $Q_{\textrm{L}}^{(0)}$, and $\Phi^{(0)}$,
assuming that the mixing terms are completely negligible up to this order.
This assumption is literally valid if $\lambdabar^2\mixingf\lesssim \epsilon^2$.
If, however, $\epsilon^2\ll \lambdabar^2\mixingf \ll \epsilon$,
the effect of mixing accumulates and becomes no longer negligible
over a propagation distance $D\sim L(\epsilon/\lambdabar^4\mixingf^2)^{1/3}$,
which is shorter than the limit of the geometric optics approximation, $L/\epsilon$,
as discussed in the previous section.
Therefore, the above conclusion is valid within the propagation distance 
$L(\epsilon/\lambdabar^4\mixingf^2)^{1/3}$ if $\epsilon^2\ll \lambdabar^2\mixingf \ll \epsilon$.
We note, however, that the estimate in Sec.~\ref{sec:Lagrangian} implies that
one has $\lambdabar^2\mixingf\ll \epsilon^2$ in a cosmological setup.

\subsection{Case (ii): \texorpdfstring{$\lambdabar^2\mixingf\gg \epsilon$}{Case (ii)}}

Next, let us consider the opposite limit, $\lambdabar^2\mixingf\gg \epsilon$.
In this regime, the three propagation eigenmodes characterized by the amplitudes
$(\widetilde{Q}_0^{(0)}, \widetilde{Q}_+^{(0)}, \widetilde{Q}_-^{(0)})$,
identified by using the leading-order equations in the previous section,
propagate at different phase velocities.

From the next-to-leading-order field equations, we obtain
\begin{align}
        &\mathcal{E}_{\mu\nu}
        -2l^\rho \delta k_\rho
        \left(Q_{\mu\nu}^{(1')}+\frac{2\alpha}{\mpl^2}i\varepsilon^{\alpha\beta\gamma}_{\;\;\;\;\;\;(\mu}
        \nabla_\alpha\phi l_\beta Q_{\nu)\gamma}^{(1')}
        \right)
        \notag \\ &
        +\frac{4\alpha}{\mpl}\left(\star C_{\mu \alpha\nu \beta}\right)
        l^\alpha l^\beta\Phi^{(1')}=0,
        \label{eq:transport-cov1}
\end{align}
and 
\begin{align}
        \mathcal{E}_\Phi+il^\mu\delta k_\mu\Phi^{(1')}
        -\frac{1}{2}\mixingf
        \left(
        e^{i\chi}
        Q_{\textrm{R}}^{(1')}-e^{-i\chi} Q_{\textrm{L}}^{(1')}
        \right)
        =0,
        \label{eq:transport-phi}
\end{align}
where $\mathcal{E}_{\mu\nu}$ and $\mathcal{E}_\Phi$ are defined by Eqs.~\eqref{def:emn} and~\eqref{def:ephi} respectively.
Note that the $\mathcal{O}(\epsilon/\mixing)$ corrections to the amplitudes
come into the transport equations of order $\mathcal{O}(\epsilon)$ through the
mixing terms of order $\mathcal{O}(\mixing)\times \mathcal{O}(\epsilon/\mixing)$.
Equation~\eqref{eq:transport-cov1} yields
\begin{align}
        \mathcal{E}_{\textrm{R}}
        +il^\mu\delta k_\mu \mathcal{A}_{\textrm{R}} Q_{\textrm{R}}^{(1')}
        +2\mixingf e^{-i\chi}\Phi^{(1')}&=0,
        \label{eq:transport-R}
        \\ 
        \mathcal{E}_{\textrm{L}}
        +il^\mu\delta k_\mu \mathcal{A}_{\textrm{L}} Q_{\textrm{L}}^{(1')}
        -2\mixingf e^{i\chi}\Phi^{(1')}&=0.
        \label{eq:transport-L}
\end{align}
For each eigenmode,
one can combine Eqs.~\eqref{eq:transport-phi}--\eqref{eq:transport-L} to eliminate
$(Q_{\textrm{R}}^{(1')}, Q_{\textrm{L}}^{(1')}, \Phi^{(1')})$ and
derive the transport equation for the corresponding amplitude.

For the mode with $l^\mu\delta k_\mu=0$ and the amplitude $\widetilde{Q}_0^{(0)}$,
Eqs.~\eqref{eq:transport-R} and~\eqref{eq:transport-L} are combined to give
\begin{align}
        e^{i\chi}\mathcal{E}_{\textrm{R}}+e^{-i\chi}\mathcal{E}_{\textrm{L}}=0,
\end{align}
with $Q_{\textrm{R}}^{(0)}=e^{-i\chi}\widetilde{Q}_0^{(0)}/\sqrt{2}$
and $Q_{\textrm{L}}^{(0)}=e^{i\chi}\widetilde{Q}_0^{(0)}/\sqrt{2}$.
This immediately leads to the transport equation,
\begin{align}
        \nabla_l\widetilde{Q}_0^{(0)}
        +\frac{1}{2}\nabla_\mu l^\mu
        \widetilde{Q}_0^{(0)}
        -\frac{i}{2}\left(\mathcal{A}_{\textrm{R}}-\mathcal{A}_{\textrm{L}}\right)\nabla_l\chi
        \widetilde{Q}_0^{(0)}
        =0.
        \label{eq:tr-tilQ0}
\end{align}
Using this transport equation we get the conservation equation,
\begin{align}
        \nabla_\mu\left(|\widetilde{Q}_0^{(0)}|^2l^\mu\right)=0.
\end{align}


For the modes with $l^\mu\delta k_\mu=\pm\delta_W$ and the amplitudes $\widetilde{Q}_\pm^{(0)}$,
Eqs.~\eqref{eq:transport-phi}--\eqref{eq:transport-L} can be combined to give
\begin{align}
        \frac{e^{i\chi}}{\mathcal{A}_{\textrm{R}}}\mathcal{E}_{\textrm{R}}
        -\frac{e^{-i\chi}}{\mathcal{A}_{\textrm{L}}}\mathcal{E}_{\textrm{L}}
        \pm 2i\sqrt{\frac{2}{\mathcal{A}_{\textrm{R}}\mathcal{A}_{\textrm{L}}}}\mathcal{E}_\Phi=0.
\end{align}
Noting Eqs.~\eqref{eq:def-amp-1}--\eqref{eq:def-amp-3},
one finds the transport equation
\begin{align}
        \nabla_l\widetilde{Q}_\pm^{(0)}
        +\frac{1}{2}\nabla_\mu l^\mu
        \widetilde{Q}_\pm^{(0)}
        +\frac{i}{4}
        \left(\mathcal{A}_{\textrm{R}}-\mathcal{A}_{\textrm{L}}\right) \nabla_l\chi \widetilde{Q}_\pm^{(0)}
        =0,
\end{align}
which results in the conservation equation
\begin{align}
        \nabla_\mu\left(|\widetilde{Q}_\pm^{(0)}|^2 l^\mu\right)=0.
\end{align}

We thus see that the current associated with each eigenmode is conserved in the $\lambdabar^2\mixingf\gg \epsilon$ regime.
For these three modes,
$\mathcal{A}_{\textrm{R}}|Q_{\textrm{R}}^{(0)}|^2l^\mu$ and $\mathcal{A}_{\textrm{L}}|Q_{\textrm{L}}^{(0)}|^2l^\mu$
are not conserved separately,
but we can show that the sum is conserved,
\begin{align}
        \nabla_\mu \left[\left(\mathcal{A}_{\textrm{R}}
        |Q_{\textrm{R}}^{(0)}|^2+\mathcal{A}_{\textrm{L}}|Q_{\textrm{L}}^{(0)}|^2\right)l^\mu\right]=0.
\end{align}

\subsection{Case (iii): \texorpdfstring{$\lambdabar^2\mixingf\sim \epsilon$}{Case (iii)}}

In the intermediate regime, $\lambdabar^2\mixingf\sim\epsilon$, the mixing terms of order $\mathcal{O}(\mixingf)$
are dropped from the leading-order equations, and the leading-order dispersion relation is null
for the three eigenmodes $(Q_{\textrm{R}}^{(0)},Q_{\textrm{L}}^{(0)},\Phi^{(0)})$.
The mixing terms are pushed into the $\mathcal{O}(\epsilon)$ transport equations, yielding
\begin{align}
        \mathcal{E}_{\mu\nu}
        +\frac{4\alpha}{\mpl}\left(\star C_{\mu \alpha\nu \beta}\right)
        l^\alpha l^\beta\Phi^{(0)}=0,
        \label{eq:transport-cov2}
\end{align}
and
\begin{align}
        \mathcal{E}_\Phi-
        \frac{\mixingf}{2}\left(
        e^{i\chi}
        Q_{\textrm{R}}^{(0)}-e^{-i\chi} Q_{\textrm{L}}^{(0)}
        \right)
        =0
        \label{eq:int-c-1}.
\end{align}
It follows from Eq.~\eqref{eq:transport-cov2} that
\begin{align}
        \mathcal{E}_{\textrm{R}}
        +2\mixingf e^{-i\chi}\Phi^{(0)}&=0,
        \label{eq:int-c-2}
        \\ 
        \mathcal{E}_{\textrm{L}}
        -2\mixingf e^{i\chi}\Phi^{(0)}
        &=0
        \label{eq:int-c-3}.
\end{align}
The transport of $(Q_{\textrm{R}}^{(0)}, Q_{\textrm{L}}^{(0)}, \Phi^{(0)})$ is governed
by the coupled equations~\eqref{eq:int-c-1}--\eqref{eq:int-c-3},
which can in general be solved numerically along the null congruence.
These equations imply
\begin{align}
        \nabla_\mu\left(\mathcal{A}_{\textrm{R}}|Q_{\textrm{R}}^{(0)}|^2 l^\mu\right)
        &=-2\mixingf\left(e^{-i\chi}\Phi^{(0)}\bar{Q}_{\textrm{R}}^{(0)}
        +\text{c.c.}\right),
        \\ 
        \nabla_\mu\left(\mathcal{A}_{\textrm{L}}|Q_{\textrm{L}}^{(0)}|^2 l^\mu\right)
        &=2\mixingf\left(e^{i\chi}\Phi^{(0)}\bar{Q}_{\textrm{L}}^{(0)}
        +\text{c.c.}\right),
        \\
        \nabla_\mu\left(|\Phi^{(0)}|^2l^\mu\right)&=
        \frac{\mixingf}{2}\Bigl(
                e^{-i\chi}\Phi^{(0)}\bar{Q}_{\textrm{R}}^{(0)}-
                e^{i\chi}\Phi^{(0)}\bar{Q}_{\textrm{L}}^{(0)}
                \notag \\ & \quad 
                +\text{c.c.}
        \Bigr),
\end{align}
showing that the individual currents are not conserved separately due to the mode mixing at the transport order,
whereas the total current
\begin{align}
        &\nabla_\mu\mathcal{J}^\mu=0,
        \\
        &\mathcal{J}^\mu \coloneq
        \left(\frac{\mathcal{A}_{\textrm{R}}}{4}|Q_{\textrm{R}}^{(0)}|^2
        +\frac{\mathcal{A}_{\textrm{L}}}{4}|Q_{\textrm{L}}^{(0)}|^2+|\Phi^{(0)}|^2\right)l^\mu,
\end{align}
is conserved.
The ghost bound $\mathcal{A}_{\textrm{R,L}}>0$ guarantees that each term in the parentheses is nonnegative.

Note that the transport equations in case (i) are reproduced by taking the limit
$\mixingf\to 0$ in the above equations.

\section{Conclusions}\label{sec:conclusion}

In this paper, we have studied the propagation of high-frequency gravitational waves
and a scalar-field fluctuation in dynamical Chern--Simons (dCS) gravity.
We have assumed a general background metric and a general background scalar-field configuration,
extending the previous work~\cite{Garfinkle:2010zx}.
Due to the nonvanishing scalar-field background, third-order derivatives appear in the linearized field equations,
which would be dominant in the naive high-frequency limit.
However, this would result in an instability
stemming from ghost-like tensor modes
even at the level of classical field theory.
To avoid this instability, we have restricted ourselves to the frequency range in which
the linearized system is free of ghosts.
The covariant ghost-free condition we have derived is given by Eq.~\eqref{eq:avoid-ghost-cov}.

Within the ghost-free frequency range, we have performed the geometric optics analysis of dCS gravity.
In addition to the usual small expansion parameter $\epsilon$ of the geometric optics approximation,
we have introduced another small quantity, $\mixingf\,(\ll \lambdabar^{-2})$,
which is expressed in terms of the CS coupling and
the Weyl tensor and characterizes tensor-scalar mixing locally.
We have distinguished three regimes: (i) $\lambdabar^2\mixingf\ll\epsilon$,
(ii) $\lambdabar^2\mixingf\gg \epsilon$, and (iii) $\lambdabar^2\mixingf\sim \epsilon$.
For these three regimes, we have determined the dispersion relations from the leading-order
equations in the $\epsilon$ expansion, and then studied the properties of the transport equations
for the leading-order amplitudes
obtained at next-to-leading order.

In case (i),
the mode mixing among the right-handed tensor mode ($Q_{\textrm{R}}^{(0)}$),
the left-handed tensor mode ($Q_{\textrm{L}}^{(0)}$), and the scalar mode ($\Phi^{(0)}$)
can be ignored even at the transport order, and
the three modes share the null dispersion relation.
The current associated with each mode is conserved separately.
The effect of the CS term appears only in the amplitude birefringence factors.
We have thus obtained a covariant description of amplitude birefringence on general backgrounds,
extending the previous results obtained in an FLRW universe~\cite{Lue:1998mq, Alexander:2004wk, Alexander:2007kv}.
Using the separate conservation of the currents,
we have shown that
the relative amplitude of the two helicity modes depends only on the
scalar-field gradients along the ray at the endpoints and is
independent of the path in between.
The propagation of gravitational waves in spacetime with vanishing Weyl tensor falls into this case.
Even if the Weyl tensor itself is nonnegligible,
gravitational waves propagating in a principal null direction of the Weyl tensor can be described by this case.
In case (ii) where mode mixing is efficient,
we identified propagating eigenmodes and showed that
there is one luminal, one subluminal, and one superluminal mode.
The current associated with each of these three eigenmodes is conserved separately.
The appearance of one subluminal and one superluminal mode has already been pointed out in
Ref.~\cite{Garfinkle:2010zx} on the backgrounds with the vanishing scalar field and Pontryagin density.
We have shown here that the conclusion is generic and remains the same for more general backgrounds.
Case (iii) lies in the intermediate regime between the above two. In this case,
the leading-order dispersion relation is null for
the right-handed and left-handed tensor modes and the scalar mode.
Mode mixing is accounted for at transport order,
and consequently, only the total current is conserved.

Throughout the paper, we have studied the structure of dCS gravity as a classical field theory.
If dCS gravity is regarded as an EFT,
within the regime of its validity only case (i) is realized
and the ghost bound is automatically satisfied.
Cases (ii) and (iii) characterize the classical theory beyond its cutoff.

We leave possible extensions of this work for future investigations.
For example, the Palatini formulation of dCS gravity~\cite{Boudet:2022wmb, Boudet:2022nub, Bombacigno:2022naf, Sulantay:2022sag}
gives rise to new gravitational-wave phenomenology that is not present in the metric formulation.
A more general framework for parity-violating (metric) scalar-tensor theories has also been developed~\cite{Crisostomi:2017ugk}.
It would be interesting to generalize the present work and
explore the consequences of covariant geometric optics in such theories.

\acknowledgments
This work was supported by JSPS KAKENHI Grant No.~JP25K07308.
The author engaged in extensive discussions with Claude Opus 5.5 (Anthropic)
throughout this work and used it to cross-check the correctness of the physical results.
All AI-assisted suggestions were critically evaluated and approved by the author.

\appendix 

\onecolumngrid

\section{Derivation of the quadratic action}\label{app:derivation-of-quadratic-action}

Here we present the explicit form of the quadratic action~\eqref{eq:I2-schematic}.

It is rather straightforward to derive the quadratic contribution from the Einstein--Hilbert term plus the scalar field:
\begin{align}
        \left(\sqrt{-g}\mathcal{L}_{\textrm{EH}+\phi}\right)^{(2)}&=
        \sqrt{-g}
        \biggl\{
        -\frac{1}{8}\nabla_{\mu}h_{\nu\rho}\nabla^\mu h^{\nu\rho}
        +\frac{1}{8}\nabla_\mu h\nabla^\mu h-\frac{1}{4}\nabla_\mu h\nabla_\nu h^{\mu\nu}
        +\frac{1}{4}\nabla_\rho h^{\mu\nu}\nabla_\mu h^{\rho}_{\nu}
        \notag \\ & \quad 
        -\frac{1}{2}(\nabla\delta\phi)^2-\frac{V_{,\phi\phi}}{2}\delta\phi^2
        -\frac{V_{,\phi}}{2\mpl}h\delta\phi
        +\frac{1}{\mpl}\nabla^\mu\phi \left(h_{\mu\nu}-\frac{1}{2}g_{\mu\nu}h\right)\nabla^\nu\delta\phi
        \notag \\ & \quad 
        +\frac{1}{2}\left(h^{\mu\rho}h_\rho^\nu-\frac{1}{2}h h^{\mu\nu}\right)R_{\mu\nu}
        -\frac{1}{8}\left(h_{\mu\nu}h^{\mu\nu}-\frac{1}{2}h^2\right)R
        \notag \\ & \quad 
        -\frac{1}{2\mpl^{2}}\left(h^{\mu\rho}h_\rho^\nu-\frac{1}{2}h h^{\mu\nu}\right)\nabla_\mu\phi\nabla_\nu\phi
        +\frac{1}{8\mpl^{2}}\left(h_{\mu\nu}h^{\mu\nu}-\frac{1}{2}h^2\right)\left[(\nabla\phi)^2+2V\right]
        \biggr\}
        .
\end{align}

The calculation of the contribution from the CS term requires some technical care.
Let us consider two different metrics $\check g_{\mu\nu}$ and $g_{\mu\nu}$.
For the moment, the former can be a perturbed metric, a conformally transformed metric, or any other metric.
One has the following relation:
\begin{align}
        \sqrt{-\check g}\check\varepsilon^{\mu\nu\rho\sigma}=
        \sqrt{-g}\varepsilon^{\mu\nu\rho\sigma}.
\end{align}
The connections for the two metrics are related as
\begin{align}
        \check \Gamma^\rho_{\mu\nu}=\Gamma^\rho_{\mu\nu}+\gamma^\rho_{\mu\nu},
\end{align}
where $\gamma^\rho_{\mu\nu}$ is some tensor.
Then, one has the exact relation
\begin{align}
        \check R_{\mu\nu\rho}^{\;\;\;\;\;\;\sigma}=R_{\mu\nu\rho}^{\;\;\;\;\;\;\sigma}
        +
        \nabla_\nu \gamma^\sigma_{\mu\rho}-\nabla_\mu \gamma^\sigma_{\nu\rho}
        +\gamma_{\mu\rho}^\alpha \gamma^\sigma_{\alpha\nu}-\gamma^\alpha_{\nu\rho}\gamma^\sigma_{\alpha\mu},
\end{align}
where $\nabla_\mu$ is a covariant derivative associated with $g_{\mu\nu}$.
Now, it is straightforward to show that
\begin{align}
        \frac{1}{4}\sqrt{-\check g}\check\varepsilon^{\mu\nu\rho\sigma}
        \check R_{\mu\nu}^{\;\;\;\;\alpha\beta}
        \check R_{\alpha\beta \rho\sigma}
        &=-\frac{1}{4}\sqrt{-g}\varepsilon^{\mu\nu\rho\sigma}
        \check R_{\mu\nu\alpha}^{\;\;\;\;\;\;\;\beta}
        \check R_{\rho\sigma\beta}^{\;\;\;\;\;\;\;\alpha}
        \notag \\ &
        =
        \frac{1}{4}\sqrt{- g} \varepsilon^{\mu\nu\rho\sigma}
        R_{\mu\nu}^{\;\;\;\;\alpha\beta}
        R_{\alpha\beta \rho\sigma}
        -\sqrt{-g}\nabla_\mu\mathcal{C}^\mu,
\end{align}
where 
\begin{align}
        \mathcal{C}^\mu\coloneq\varepsilon^{\mu\nu\rho\sigma}
                \left(
                        \gamma^\alpha_{\nu \beta}
                        \nabla_\rho \gamma^\beta_{\sigma\alpha}
                        +\frac{2}{3}\gamma^\alpha_{\nu\beta}\gamma^\beta_{\rho\gamma}\gamma^\gamma_{\sigma\alpha}
                        -\gamma_{\nu\beta}^\alpha R_{\rho\sigma\alpha}^{\;\;\;\;\;\;\;\beta}
                \right).
\end{align}

In the case of the metric perturbations, we replace $\check g_{\mu\nu}$ with
$g^{\textrm{(tot)}}_{\mu\nu}= g_{\mu\nu}+\mpl^{-1}h_{\mu\nu}$.
To second order in $h_{\mu\nu}$, we have
\begin{align}
        \gamma_{\mu\nu}^\rho=\frac{1}{2}\mpl^{-1}\left(g^{\rho\alpha}-\mpl^{-1}h^{\rho\alpha}\right)
        \left(
                \nabla_\mu h_{\nu\alpha}+\nabla_\nu h_{\mu\alpha}-\nabla_\alpha h_{\mu\nu}
        \right),
\end{align}
leading to $\mathcal{C}^\mu=\mpl^{-1}\mathcal{C}^\mu_{(1)}+\mpl^{-2}\mathcal{C}^\mu_{(2)}$ with
\begin{align}
        \mathcal{C}^\mu_{(1)}&=-\varepsilon^{\mu\nu\rho\sigma}R_{\rho\sigma\alpha}^{\;\;\;\;\;\;\beta}
        \nabla_\beta h_\nu^\alpha
        ,
        \\ 
        \mathcal{C}^\mu_{(2)}&=
        \frac{1}{4}\varepsilon^{\mu\nu\rho\sigma}
        \left(
                -R_{\rho\sigma\alpha}^{\;\;\;\;\;\;\beta}
                h^{\alpha\gamma}\nabla_\nu h_{\beta\gamma}
                +2\nabla_\beta h_{\nu\alpha}\nabla_\rho\nabla^\alpha h_\sigma^\beta
                -2\nabla_\beta h_{\nu\alpha}\nabla_\rho\nabla^\beta h_\sigma^\alpha
        \right)
        \notag \\ & \quad 
        +\frac{1}{2}\varepsilon^{\mu\nu\rho\sigma}
        h^{\alpha\gamma}\left(
                \nabla_\nu h_{\beta\gamma}+\nabla_\beta h_{\nu\gamma}
                -\nabla_\gamma h_{\nu\beta}
        \right)R_{\rho\sigma\alpha}^{\;\;\;\;\;\;\beta}
        .
\end{align}
Performing integration by parts, we obtain the quadratic contribution from the CS term as
\begin{align}
        \alpha^{-1}\left(\sqrt{-g}\mathcal{L}_{\textrm{CS}}\right)^{(2)}&=
        \sqrt{-g}\biggl[
        \frac{1}{4\mpl^2}\varepsilon^{\mu\nu\rho\sigma}\nabla_\mu\phi
        \left(\nabla_\beta h_{\nu\alpha}\nabla_\rho\nabla^\alpha h_\sigma^\beta
        -\nabla_\beta h_{\nu\alpha}\nabla_\rho\nabla^\beta h_\sigma^\alpha\right)
        \notag \\ & \quad 
        +\frac{1}{\mpl}\left(\star R^{\mu\nu\alpha\beta}\right)
        \nabla_\mu\delta\phi\nabla_\alpha h_{\beta\nu}
        \notag \\ & \quad 
        +\frac{1}{2\mpl^2}\nabla_\mu\phi
        \left(\star R^{\mu\nu\alpha\beta}\right)
        h_{\alpha}^{\gamma}
        \left(
                \frac{1}{2}\nabla_\nu h_{\beta\gamma}+\nabla_\beta h_{\nu\gamma}
                -\nabla_\gamma h_{\nu\beta}
        \right)
        \biggr].
\end{align}

\twocolumngrid

\bibliography{refs}
\bibliographystyle{JHEP}
\end{document}